\documentclass[doublecol]{epl2}

\usepackage{amsmath}
\usepackage{amsfonts}

\newcommand{\fig}[1]{fig.\,\ref{#1}}
\newcommand{\eq}[1]{eq.\,(\ref{#1})}

\newcommand{\dd}{\text{d}}

\newcommand{\ebold}{\textbf{e}}

\newcommand{\rbold}{\textbf{r}}

\newcommand{\Abold}{\textbf{A}}

\newcommand{\Kbold}{\textbf{K}}

\newcommand{\Rbold}{\textbf{R}}

\newcommand{\I}{\text{i}}

\newcommand{\E}{\text{e}}

\newcommand{\hc}{\text{hc}}

\DeclareMathAlphabet\mathbfcal{OMS}{cmsy}{b}{n}

\newcommand{\xunitvect}{\mathbf{e}_x}
\newcommand{\yunitvect}{\mathbf{e}_y}

\usepackage[caption=false]{subfig}
\newcommand{\phantomsubfloat}[1]{
    {% apply caption setup only temporarily
        \captionsetup[subfigure]{labelformat=empty}
        \subfloat[][]{#1}
    }%
}
\newcommand{\captionlabel}[1]{(\protect\subref{#1})}

\renewcommand{\figurename}{Fig.}

\begin{document}

\title{
Atomic topological quantum matter using synthetic dimensions
}
\author{A. Fabre\inst{1,2,3} \and S. Nascimbene\inst{1}}
\shortauthor{A. Fabre and S. Nascimbene}

\institute{
  \inst{1} Laboratoire Kastler Brossel,  Coll\`ege de France, CNRS, ENS-PSL University, Sorbonne Universit\'e, 11 Place Marcelin Berthelot, 75005 Paris, France\\
  \inst{2} Institute of Physics, Ecole Polytechnique F\'ed\'erale de Lausanne, CH-1015 Lausanne, Switzerland\\
  \inst{3} Center for Quantum Science and Engineering, Ecole Polytechnique F\'ed\'erale de Lausanne, CH-1015 Lausanne, Switzerland
}

\abstract{
The realization of topological states of matter in ultracold atomic gases is currently the subject of intense experimental activity. Using a synthetic dimension, encoded in a non-spatial degree of freedom, can greatly simplify the  simulation of gauge fields and give access to exotic topological states. We review here recent advances in the field and discuss future perspectives for interacting systems.
}
 
\maketitle

\begin{figure}[t!]
\begin{center}
 \includegraphics[
 draft=false,scale=1.2,
 trim={0mm 4mm 0 0mm},
 ]{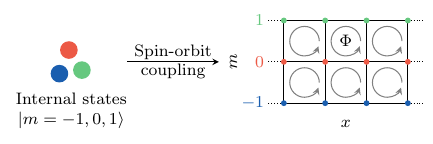}
 \end{center}
\caption{
Scheme of an atomic system whose dynamics combines one spatial dimension $x$ and one synthetic dimension $m$, encoded in internal spin states on this example\cite{mancini_observation_2015,stuhl_visualizing_2015}. A spin-orbit coupling gives rise to an artificial gauge field in this (1+1)-dimensional space with uniform flux $\Phi$, resulting in topological bands. 
\label{fig:synthetic}}
\end{figure}

In recent years, there has been a significant amount of research focused on topological systems, motivated by their unique physical properties and the potential applications  in quantum technologies \cite{hasan_colloquium_2010}. They are characterized by a non-trivial topological invariant, which describes a physical property  that remains invariant under weak deformations.
Electronic systems exhibiting a quantum Hall effect were the first examples of non-trivial topological states. Soon after their discovery  \cite{klitzing_new_1980}, it was realized that the  robust quantization of the Hall conductance  as a multiple of $e^2/h$ is a direct manifestation of a non-zero Chern number that characterizes the topology of Bloch bands \cite{thouless_quantized_1982}.   The discovery of the quantum spin Hall effect \cite{konig_quantum_2007} and topological insulators \cite{hsieh_topological_2008} introduced new examples of topological systems with distinct  topological invariants. Over time, a wide range of topological systems has been predicted and classified based on their  dimensionality and underlying symmetries \cite{chiu_classification_2016}. While some of these systems have been found in solid-state materials \cite{bansil_colloquium_2016}, artificial systems such as solid-state heterostructures, quantum circuits, photonic or atomic systems offer interesting alternatives to explore the variety of topological phases of matter \cite{aidelsburger_artificial_2018}. Synthetic dimensions, i.e. the encoding of a dimension in a non-spatial degree of freedom (see \fig{fig:synthetic}), can provide an efficient way to generate exotic topological systems \cite{celi_synthetic_2014}.  In this review, we focus on the possibilities offered by synthetic dimensions for simulating topological phases using ultracold atomic systems. This review does not cover photonic systems, in which synthetic dimensions have been used extensively \cite{yuan_synthetic_2018,lustig_topological_2021}. For a broader perspective on the use of synthetic dimensions across different physical platforms, we recommend the review by Ozawa et al.~\cite{ozawa_topological_2019}.

\section{Simulating gauge fields in atomic gases}

Topological phases of matter arise when particles are subjected to an external gauge field, typically a spin-orbit coupling or an orbital magnetic field \cite{hasan_colloquium_2010}. 
The action of an effective magnetic field $B$ on a particle of charge $q$ can be described by the Aharonov-Bohm phase $\Phi_{\mathcal{C}} = \frac{q}{\hbar} \int_\mathcal{C} \Abold(\rbold) \, \dd \rbold$ that the particle acquires when evolving adiabatically along a path $\mathcal{C}$. Unlike the usual dynamical phase, the Aharonov-Bohm phase only depends on the path geometry and the vector potential $\Abold(\rbold)$.

\begin{figure*}[ht!]
 \includegraphics[
 draft=false,scale=0.98,
 trim={0mm 0mm 0mm 0mm},
 clip
 ]{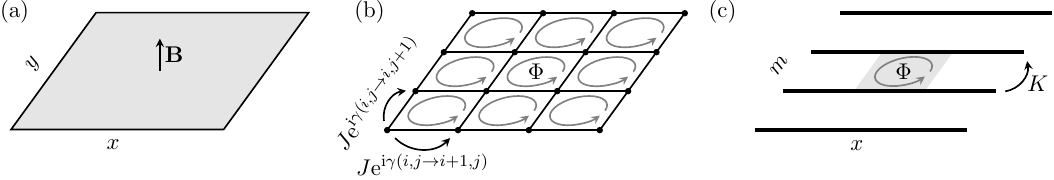}
    \phantomsubfloat{\label{fig:model_LLL}}
    \phantomsubfloat{\label{fig:model_HH}}
    \phantomsubfloat{\label{fig:model_Kane}}
\vspace{-9mm}
\caption{
Models of charged particle dynamics in the presence of a magnetic field. 
\captionlabel{fig:model_LLL} Continuous Hall system in two dimensions in the presence of an orbital magnetic field. 
\captionlabel{fig:model_HH} Harper-Hofstadter model on a discrete lattice, in which complex-hoppings give rise to a non-zero magnetic flux $\Phi$ per plaquette.
\captionlabel{fig:model_Kane} Coupled wire model, representing a hybrid case with a continuous dimension $x$ and a discrete dimension $m$. 
\label{fig:models}}
\end{figure*}

For neutral atoms, applying a magnetic field does not give rise to an Aharonov-Bohm phase. Alternatively, gauge fields can be artificially produced  by engineering different kinds of geometrical phases, namely the Sagnac phase for rotating gases, the Berry phase for laser-dressed systems, or the Peierls phase in discrete lattice systems. We refer to the review \cite{dalibard_colloquium_2011} for a comprehensive description of the different protocols.

\section{Iconic models of topological systems}
The effects of these geometrical phases can generally be recast in a collection of models describing topological systems. In this section, we focus on several models that are specifically relevant for synthetic dimensions (see \fig{fig:models}).

\subsection{Landau levels}
We begin with the Landau Hamiltonian, which describes the motion of a charged particle  in an $xy$ plane under the influence of a perpendicular magnetic field $B$ (see \fig{fig:model_LLL}).   It is given by
\begin{equation*}
H=\frac{(p_x-eBy)^2}{2M}+\frac{p_y^2}{2M},
\end{equation*}
where we assume a particle of charge $q=-e$, and we express the vector potential  in the Landau gauge as $\Abold=-By\,\ebold_x$.

Beyond  electronic gases subjected to a magnetic field, the Landau Hamiltonian is relevant for the description of continuous atomic systems exhibiting a Hall effect, such as rotating or laser-dressed atomic gases \cite{madison_vortex_2000,abo-shaeer_observation_2001,bretin_fast_2004,schweikhard_rapidly_2004,lin_synthetic_2009}.

\subsection{The Harper-Hofstadter model}
In lattice systems, topological bands can also arise when the hopping of particles between neighboring lattice sites is described by complex amplitudes (see \fig{fig:model_HH}). Specifically, the complex phase involved in the hopping process from $\rbold_1$ to $\rbold_2$, known as the Peierls phase, can be interpreted as an integrated Aharonov-Bohm phase $\gamma(\rbold_1 \rightarrow \rbold_2) = q/\hbar \int^{\rbold_2}_{\rbold_1} \Abold(\rbold) \dd \rbold$. 
The Harper-Hofstadter model describes  a square lattice model with such Peierls phases, as~\cite{harper_single_1955,hofstadter_energy_1976}
\begin{equation}
\label{eq:HH}
\begin{aligned}
    H = -J\sum_{ i,j} &~ \E^{\I \gamma(i,j \rightarrow i+1,j)} \hat{a}_{i+1,j}^\dagger \hat{a}_{i,j} \\
    &+  \E^{\I \gamma(i,j \rightarrow i,j+1)} \hat{a}_{i,j+1}^\dagger \hat{a}_{i,j} + \hc,
\end{aligned}
\end{equation}
where the operator $\hat{a}_{i,j}$ annihilates a particle at position $\Rbold = ia \xunitvect + ja \yunitvect$ (with integers $i$, $j$), where $a$ is the lattice spacing. The  Peierls phases are associated with a magnetic flux  $\Phi=a^2 B$ through a unit cell. This model exhibits magnetic Bloch bands with a non-trivial topological character described by an integer topological invariant, the Chern number \cite{thouless_quantized_1982}.

The Harper-Hofstadter model, along with other lattice models that possess non-zero Chern numbers like the Haldane model, have been successfully realized using ultracold atoms in optical lattices. This has been achieved by imprinting complex phases on the tunnel hoppings through techniques such as resonant shaking or light-assisted processes \cite{aidelsburger_realization_2013,miyake_realizing_2013,jotzu_experimental_2014,eckardt_colloquium_2017}. The topological properties of these bands have been experimentally demonstrated through various methods, including the measurement of Berry curvature \cite{flaschner_experimental_2016} and the direct observation of transverse response quantization~\cite{aidelsburger_measuring_2015}. A microscopic version of this model has also been realized in Rydberg atom arrays \cite{lienhard_realization_2020}.

\subsection{The coupled-wire model} The model examined by Kane et al \cite{kane_fractional_2002} explores a unique scenario where one dimension is continuous and the other is discrete.  It  thus consists of  an array of coupled (one-dimensional) quantum wires (see \fig{fig:model_Kane}), which allows one to use  theoretical tools from 1D physics, such as Luttinger liquid theory and bosonization,  for the description of strongly interacting topological states in two dimensions~\cite{meng_coupled-wire_2020}. 

In the Landau gauge $\Abold=-B y \,\ebold_x$, the coupled wire Hamiltonian writes
\begin{equation}
\begin{aligned}
    H  = \sum_{m,p_x} &\frac{(p_x + qB a m)^2}{2M} \hat{a}_{m,p_x}^\dagger \hat{a}_{m,p_x}  \\
    & - K (\hat{a}_{m+1,p_x}^\dagger \hat{a}_{m,p_x} + \hc),
\end{aligned}\label{eq:H_Kane}
\end{equation}
where $a$ is the distance between adjacent wires,  the operator $\hat{a}_{m,p_x}$ annihilates a particle with momentum $p_x$ in wire $m$ (located at position $m a$ along $y$, with integer $m$), and we introduce the coupling strength $K$ between neighouring wires.

\section{Synthetic dimensions}  The concept of synthetic dimension was initially introduced by Boada and Celi et al \cite{boada_quantum_2012,celi_synthetic_2014} as a way to encode one or more dimensions in a non-spatial degree of freedom. It encompasses various approaches depending on the nature of the synthetic dimension. In the following, we provide different examples that utilize the internal spin, momentum states, or time-modulation quanta. Synthetic dimensions offer several advantages for quantum simulation, including simplified generation of gauge fields, programmability of dimensionality through connectivity between sites, simulation of physics in dimensions higher than three ($D>3$), and the presence of sharp boundaries (\fig{fig:synthetic}).

\subsection{Internal spin}

\begin{figure}[t!]
\begin{center}
 \includegraphics[
 draft=false,scale=0.92, %0.88
 trim={0mm 0mm 0 0mm},
 ]{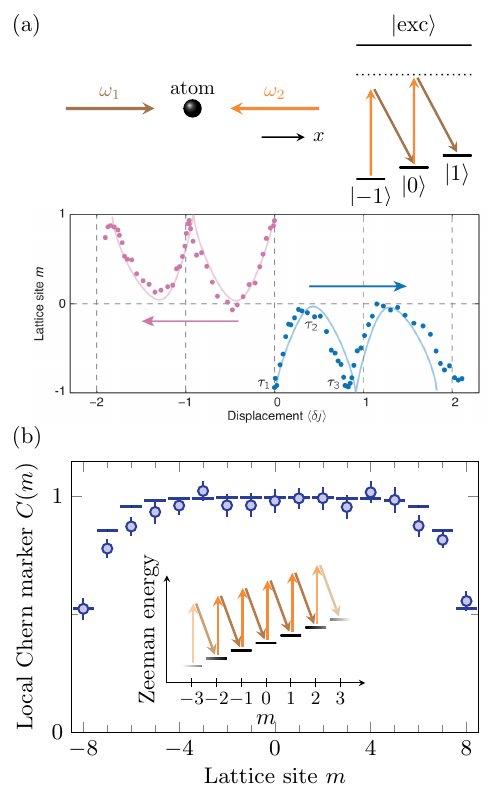}
 \end{center}
    \phantomsubfloat{\label{fig:spin:smallspin}}
    \phantomsubfloat{\label{fig:spin:largespin}}
\vspace{-13mm}
\caption{
\captionlabel{fig:spin:smallspin}  Top: Laser scheme used to create a spin-orbit coupling between spin projection states $m$ and motion along $x$. Bottom: Time evolution of the center-of-mass, showing two skipping orbits of opposite chirality bouncing at the two edges $m=\pm1$ of the synthetic dimension (from \cite{stuhl_visualizing_2015}). \captionlabel{fig:spin:largespin} Local Chern marker quantifying the local Hall response as a function of $m$. The large spin $J=8$ of dysprosium allows defining a wide bulk region over which the Hall response is quantized (adapted from \cite{chalopin_probing_2020}).
\label{fig:spin}}
\end{figure}

\begin{figure*}[t!]
 \includegraphics[
 draft=false,scale=1.,
 trim={0mm 0mm 0 0mm},
 ]{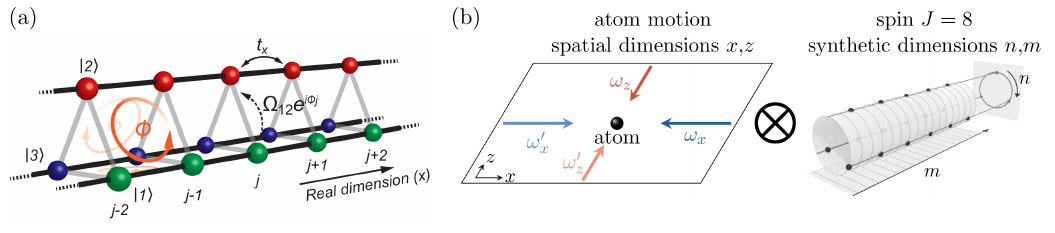}
    \phantomsubfloat{\label{fig:exotic:cylinder}}
    \phantomsubfloat{\label{fig:exotic:4D}}
\vspace{-10mm}
\caption{
\captionlabel{fig:exotic:cylinder}  Quantum Hall cylinder extended along $x$, with three cyclically-coupled internal states \cite{han_band_2019}. \captionlabel{fig:exotic:4D} Four-dimensional quantum Hall system by coupling the atomic motion along two dimensions (the third spatial dimension remaining decoupled) to the spin $J=8$ of dysprosium atoms, in which two  synthetic dimensions are encoded \cite{bouhiron_realization_2022}.
\label{fig:exotic}}
\end{figure*}

The initial implementations of a synthetic dimension involved creating a light-induced spin-orbit coupling between the atomic motion along a spatial dimension $x$ and a set of three internal spin states labeled by an integer $m\in\{-1,0,1\}$ \cite{mancini_observation_2015,stuhl_visualizing_2015}. This approach drew inspiration from previous work by the Spielman group on generating artificial gauge fields for neutral atoms~\cite{juzeliunas_light-induced_2006,lin_synthetic_2009}. By using a two-photon Raman process with counter-propagating laser beams of momentum $k$, a spin transition $m\rightarrow m+1$ can be induced, imparting a momentum kick $-2\hbar k\,\ebold_x$ to the atoms (see \fig{fig:spin:smallspin}). The conserved canonical momentum $p_x$ is defined as $p_x=Mv_x+2\hbar k m$, where $M$ is the mass of the atoms. The single-particle Hamiltonian for this system can be expressed as
\begin{equation*}
H=\!\sum_{m,p_x}\!\frac{(p_x-2\hbar k m)^2}{2M} \hat{a}_{m, p_x}^\dagger \hat{a}_{m, p_x}-K(\hat{a}_{m+1, p_x}^\dagger \hat{a}_{m, p_x}\!+\hc),
\end{equation*}
which is equivalent to the coupled-wire Hamiltonian described in \eq{eq:H_Kane}, with the identification $2\hbar k m= -qBa y$.

In the initial studies of synthetic dimensions, an optical lattice with a spatial period $d$ was applied along the $x$ direction in the tight-binding regime, resulting in a discrete lattice model given by
\begin{equation*}
H=-\sum_{j,m}J \E^{-\I 2k d j}\hat{a}_{j+1,m}^\dagger \hat{a}_{j,m}+K\hat{a}_{j,m+1}^\dagger \hat{a}_{j,m},
\end{equation*}
where $J$ represents the tunnel coupling along $x$. This Hamiltonian corresponds to the Harper-Hofstadter model given in \eq{eq:HH} upon the identification $2kd=qa^2B/\hbar$.

The absence of couplings between the extremal states $m=\pm1$ results in a ribbon geometry with sharp edges, which led to the striking observation of edge modes with chiral dynamics (see \fig{fig:spin:smallspin}) \cite{mancini_observation_2015,stuhl_visualizing_2015}. These edge modes are directly connected to a non-trivial band topology, owing to the bulk-boundary correspondence. It is worth noting that the observation of topological edge modes in systems involving only spatial dimensions has been achieved only recently  \cite{braun_real-space_2023,yao_observation_2023}, highlighting the advantage of synthetic dimensions in exploring topological edge physics.
By extending this protocol to larger spins, such as $J=8$ for dysprosium or $J=6$ for erbium atoms, it becomes possible to define a broader synthetic dimension where the bulk region exhibits quantized Hall conductance ~\cite{chalopin_probing_2020,roell_chiral_2023} (see \fig{fig:spin:largespin}).

The use of synthetic dimensions based on internal states also allows for the engineering of more complex geometries beyond planar ones~\cite{boada_quantum_2015}. Cyclic couplings in the synthetic dimension gives rise to periodic boundary conditions, which has been used to generate quantum Hall systems in a cylindrical geometry  \cite{yan_emergent_2019,han_band_2019,anderson_realization_2020,luo_tunable_2020,liang_coherence_2021,fabre_laughlins_2022,li_bose-einstein_2022,burba_topological_2023}. In such systems, the absence of edges leads to fully gapped bands characterized by well-defined topological invariants  (see \fig{fig:exotic:cylinder}).

By employing more intricate spin coupling configurations, it is possible to simulate more than a single synthetic dimension \cite{fabre_simulating_2022}. This advancement has led to the realization of a four-dimensional quantum Hall system by combining two spatial dimensions with two synthetic dimensions encoded in the large electronic spin of dysprosium atoms \cite{bouhiron_realization_2022} (see \fig{fig:exotic:4D}). Furthermore, implementing inhomogeneous spin coupling amplitudes can simulate dynamics with a position-dependent mass, which is particularly relevant for the simulation of exotic Hamiltonians, such as the entanglement Hamiltonian of a quantum Hall system \cite{redon_realizing_2023}.

In addition to quantum gases composed of atoms in their electronic ground states, polar molecules and Rydberg atoms have emerged as promising candidates for implementing large synthetic dimensions. These dimensions can be encoded either in the rotational degree of freedom of the molecules~\cite{sundar_synthetic_2018,sundar_strings_2019,blackmore_coherent_2020} or in the level structure of Rydberg atoms~\cite{hummel_synthetic_2021,kanungo_realizing_2022}.
\begin{figure}[b!]
\begin{center}
 \includegraphics[
 draft=false,scale=0.98,
 trim={0mm 0mm 2mm 0mm},
 clip
 ]{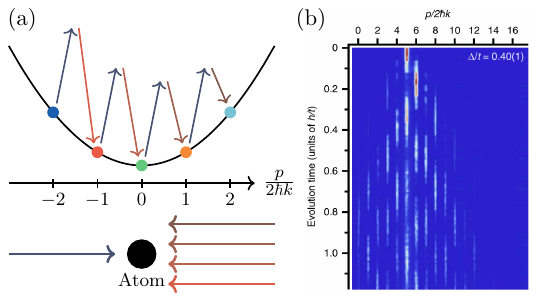}
 \end{center}
    \phantomsubfloat{\label{fig:momentum:scheme}}
    \phantomsubfloat{\label{fig:momentum:SSH}}
\vspace{-12mm}
\caption{
\captionlabel{fig:momentum:scheme}   Two-photon transition scheme used to couple a discrete set of momentum states using a multi-tone laser. 
\captionlabel{fig:momentum:SSH} Dynamics of a wavepacket prepared close to the edge of an SSH chain. The observed slow dynamics signals the population of a localized topological edge mode (from \cite{meier_observation_2016}).
\label{fig:momentum}}
\end{figure}

\subsection{Momentum states and other external degrees of freedom}

Synthetic dimensions can also be implemented using an arbitrary set of quantum states associated with the motional degree of freedom. In this context, we will focus on the utilization of momentum states, which has seen significant development. The synthetic dimension is established by employing a pair of lasers that induce two-photon Bragg scattering. Each transition results in an exchange of momentum between the light and atoms, which is quantized by the momentum difference between the two lasers, denoted as  $\Kbold=K\,\ebold_x$. Assuming an initial momentum $p_0$ along $x$, the atom's momentum explores a discrete set of momenta  $p_x=p_0 + n \hbar K$, with $n$ integer, which can be used to define a  synthetic dimension  \cite{meier_atom-optics_2016}.

Due  to the quadratic dispersion of the kinetic energy, each optical transition needs to be individually addressed to ensure resonant coupling. A common scheme relies on a beam with a career frequency and a multi-tone beam, such that each tone associated with the career resonantly couples neighbouring momentum states (see \fig{fig:momentum:scheme}). This scheme allows for full programmability, enabling individual addressing of the complex-valued hopping matrix elements between momentum states. With this protocol, various lattice models have been successfully simulated, \cite{meier_atom-optics_2016}, including the topological SSH model \cite{meier_observation_2016,xie_topological_2020,yuan_realizing_2023} (see \fig{fig:momentum:SSH}), disordered systems \cite{an_engineering_2018,meier_observation_2018,xiao_observation_2021}, models with dissipation \cite{lapp_engineering_2019,gou_tunable_2020,liang_dynamic_2022} and chaotic dynamics \cite{meier_exploring_2019}. Topological flux lattices could also be engineered by generalizing the scheme to two dimensions \cite{an_direct_2017}.

Other motional degrees of freedom can also be used to define synthetic dimensions. Orbitals states of an optical lattice have been employed to realize  different types of ladder systems~\cite{li_topological_2013,kang_realization_2018}, and a recent study has explored the use of eigenmodes of a harmonic trapping potential~\cite{price_synthetic_2017,salerno_quantized_2019,oliver_bloch_2023}.

\subsection{Floquet-engineered synthetic dimensions}

\begin{figure}[t!]
\begin{center}
 \includegraphics[
 draft=false,width=6.8cm,trim= 0 15 0 0
 ]{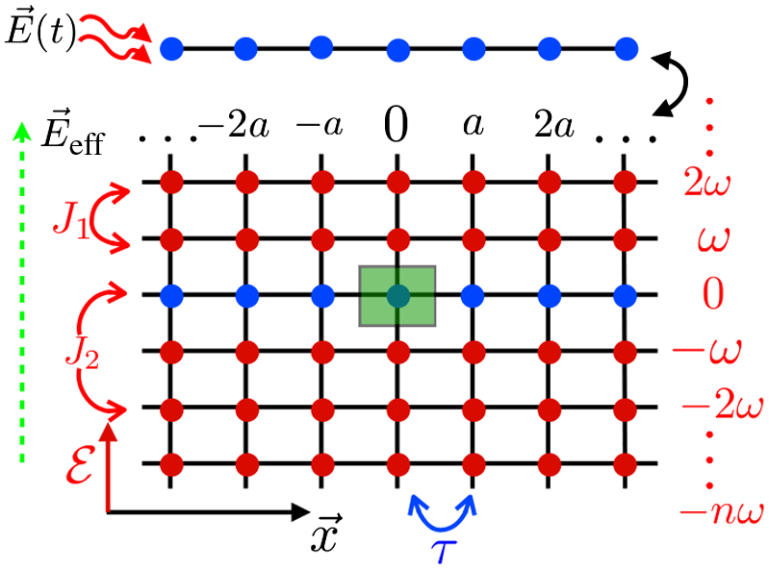}
 \end{center}
\caption{
Scheme of a 1D chain subjected to a time-periodic modulation of frequency $\omega$. The quanta of energy $n\hbar\omega$ absorbed by the system can be used to define a synthetic dimension (from \cite{gomez-leon_floquet-bloch_2013}).
\label{fig:floquet}}
\end{figure}

Time modulation, often referred to as Floquet engineering, has proven to be a successful approach for realizing complex quantum systems  \cite{goldman_periodically_2014,bukov_universal_2015,eckardt_colloquium_2017,rudner_band_2020}.  A notable example of its application is the quantum kicked rotor, which serves as a powerful tool for simulating disordered quantum systems \cite{moore_atom_1995,chabe_experimental_2008}. Time-modulated systems can give rise to topological physics in various contexts. In the regime of low-frequency driving, adiabatic cycles can lead to a quantized particle transport phenomenon known as the topological Thouless pump \cite{thouless_quantization_1983,nakajima_topological_2016,lohse_thouless_2016,citro_thouless_2023}. On the other hand, in the high-frequency modulation regime, it is possible to define effective static Hamiltonians that break time-reversal symmetry and exhibit topological Bloch bands \cite{aidelsburger_realization_2013,miyake_realizing_2013,jotzu_experimental_2014,flaschner_experimental_2016,eckardt_colloquium_2017}. 

The most complex situation takes place at intermediate modulation frequency, where  unique topological phases can be generated that have no equivalent in static systems  \cite{rudner_anomalous_2013,wintersperger_realization_2020}.  
In this regime, the number of energy quanta $\hbar\omega$ absorbed by the system from the driving field can act as a synthetic dimension \cite{gomez-leon_floquet-bloch_2013} (see \fig{fig:floquet}). This scheme has been implemented in an atomic kicked rotor system, allowing to simulate disordered systems in the presence of a gauge field \cite{hainaut_controlling_2018}.  More generally, this concept has yielded significant advancements in the field of photonics \cite{yuan_synthetic_2018,lustig_topological_2021} and holds great potential for implementation in ultracold atomic gases.

\section{Perspectives: interactions in systems with synthetic dimensions}

\begin{figure}[t!]
\begin{center}
 \includegraphics[
 draft=false,scale=0.95,
 trim={0mm 0mm 0 0mm},
 ]{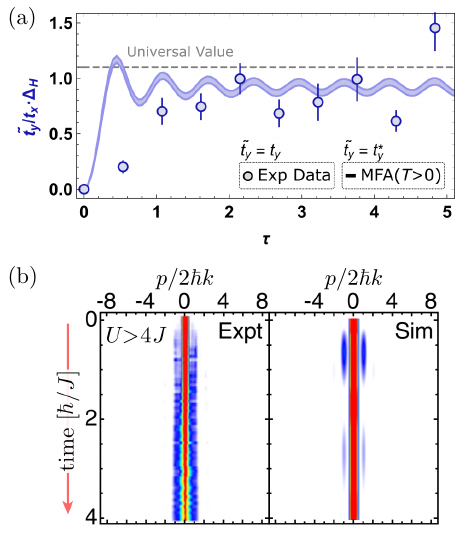}
 \end{center}
    \phantomsubfloat{\label{fig:interactions:ladder}}
    \phantomsubfloat{\label{fig:interactions:self-trapping}}
\vspace{-9mm}
\caption{
\captionlabel{fig:interactions:ladder}  Time evolution of the Hall imbalance characterizing the Hall response in a ladder system. It reaches a plateau of universal value, independent of the interaction strength (from \cite{zhou_observation_2023}).
\captionlabel{fig:interactions:self-trapping}  Macroscopic self-trapping in an array of laser-coupled momentum states (from \cite{an_nonlinear_2021}).
\label{fig:interactions}}
\end{figure}

Although the classification of topological states of matter has been achieved in the absence of interparticle interactions, the relationship between interactions and topology is still not well-understood. Drawing an analogy to the fractional quantum Hall effect, it is anticipated that when interacting particles evolve in topological Bloch bands, they may give rise to intricate states of matter with strong correlations, characterized by a topological order. 

A first advantage of synthetic dimensions is the significant energy gap that protects the topological properties of the ground band \cite{chalopin_probing_2020,roell_chiral_2023}, as opposed to conventional lattice systems. This increase in energy scales could be valuable for creating many-body states that require low entropy.

When employing synthetic dimensions to engineer topological Bloch bands, the unique nature of interactions can give rise to specific phenomena. We first consider the case of synthetic dimensions encoded in the internal spin. While interactions are typically short-ranged in quantum gases, two particles a priori interact for any combination of spin states, so that the interaction range is  infinite along synthetic dimensions. An important special case occurs when a  synthetic dimension is encoded in the nuclear spin of two-electron atoms. In this situation, the interaction becomes decoupled from the spin, exhibiting SU($N$) symmetry \cite{cazalilla_ultracold_2014}. This means that the interaction strength remains uniform irrespective of the distance along the synthetic dimension. 

In hybrid systems with both spatial and synthetic dimensions, interactions are strongly anisotropic, with a short (resp. long) range along spatial (resp. synthetic) dimensions. In ladder geometries with a few sites along the synthetic dimension, theoretical studies have predicted the existence of various phases with different charge and spin order (see \cite{ozawa_topological_2019} for a  review). Recently, a minimal system using two spin states has been studied experimentally \cite{zhou_observation_2023}, demonstrating a universal Hall response in the strongly-interacting regime (see \fig{fig:interactions:ladder}). The conditions for the occurrence of genuine two-dimensional fractional Hall states in systems with synthetic dimensions remains to be elucidated. A possible requirement could involve reducing the interaction range,  by spatially separating the synthetic lattice sites \cite{chalopin_probing_2020} or using an interaction Trotterization scheme \cite{barbiero_bose-hubbard_2020}.

In the case of synthetic dimensions encoded in external degrees of freedom, the nature of interactions is determined by the spatial overlap between distinct eigenmodes. When utilizing momentum states, interactions have an infinite range, while the use of harmonic oscillator states leads to an algebraic decay of interactions with distance~\cite{price_synthetic_2017}. A first study of interacting Bose-Einstein condensates in a momentum-state lattice has been reported, uncovering phenomena such as self-trapping and Josephson junction physics~\cite{an_nonlinear_2021} (see \fig{fig:interactions:self-trapping}).

The field of interacting systems with synthetic dimensions is still in its early stages. Given the specific character of interactions in those systems, we anticipate the possibility to uncover a wide range of novel physical phenomena. % and many open questions to explore.

\acknowledgments
We  thank the members of the Bose-Einstein condensate team at LKB for fruitful discussions, and Raphael Lopes for careful reading of the manuscript. SN acknowledges support from European Union (grant TOPODY 756722 from the European Research Council) and Institut Universitaire de France. AF acknowledges support from the Center for Quantum Science and Engineering at EPFL.

\bibliographystyle{eplbib.bst}
%\bibliography{2023_Perspective_EPL.bib}

\end{document}